# An Interaction Sensitivity Framework in Perturbation Space for the Converter-Driven Stability of Power Systems

Jihun Kook and Jung-Wook Park*

School of Electrical and Electronic Engineering, Yonsei University, Seoul 03722, South Korea

*Corresponding author. E-mail: jungpark@yonsei.ac.kr

Contributing author: jhkook154@yonsei.ac.kr

**Abstract**

**As the integration of renewables accelerates the decarbonization process in the energy sector, power systems are becoming increasingly dominated by converter-interfaced generation. However, this results in the emergence of multiple oscillations, which can trigger cascading disconnections and potentially damage generators. To address these critical issues, modern power systems commonly employ classic participation factor (PF) analyses, which relate oscillatory modes (eigenvalues) to controller state variables. While these methods are effective, a critical yet previously unexplored challenge involves capturing the interaction reconfigurations observed after a perturbation occurs. Because PF analyses are limited to fixed operating points, they are structurally unable to analyze these event-dependent mode–state interactions under perturbations. Here, we establish an interaction sensitivity framework that is formulated in perturbation space, uncovering the mechanisms by which modes and states respond to perturbations. This framework provides analytic expressions and physically interpretable mechanisms that explain why specific control adjustments enhance or reduce stability, revealing causal relationships beyond conventional stability analysis. More broadly, this formulation provides a general analytical basis for understanding perturbation-driven interaction reconfigurations in high-dimensional dynamical networks described by state-space equations, and supports advanced controller design in complex systems.**

## Introduction

Global efforts to reduce greenhouse gas emissions are driving a rapid transition in the energy sector[1-3]. In power systems, coal- and other fossil fuel-based synchronous generators (SGs) are gradually being replaced by converter-interfaced renewable energy sources (RESs) such as wind and photovoltaic generators[3-6]. While this transition is essential for achieving sustainable energy and mitigating climate impacts, it also introduces stability issues that are not observed in conventional SG-dominated power systems[7]. Recent large-scale disturbance events, such as the total blackouts in Spain and Portugal, have highlighted the growing concerns regarding the oscillatory phenomena and dynamic interactions occurring in power systems operating with very high converter-interfaced RES generation shares[8]. These challenges arise from the fact that such systems rely on high-dimensional control structures whose dynamics are influenced by both SGs and RESs. As a result, understanding the interactions between oscillatory modes and controller states has become crucial for interpreting system behaviors and maintaining stable operations. Participation factor (PF) analysis is a well-established tool for examining these interactions, and it has been widely used to identify the sources of oscillations and to guide controller tuning processes[9-16]. However, because the PFs are computed from a system matrix that is linearized at a fixed operating point (see Supplementary Information Section 1 for a derivation of the system matrix and Methods for a derivation of the PFs), they cannot capture how mode–state relationships vary under small perturbations. Therefore, PF analyses are insufficient for describing interaction reconfigurations, underscoring the need for analytical tools that provide the physically interpretable mechanisms underlying perturbation-driven dynamics in converter-dominated power systems.

A fundamental difficulty encountered during a PF analysis arises from its formulation within a fixed space defined by the eigensolutions and PFs obtained from a system matrix that is linearized at a given operating point. This formulation assumes simple eigenvalues (see Supplementary Information Section 2 for the eigenvalue taxonomy), even though repeated eigenvalues commonly occur in converter-dominated power systems as a consequence of the controller-design-induced linear dependencies of the system matrix[17-20], fixed initial controller conditions[16,21-23], structural symmetries[24,25], and modal resonance[26-29]. We show that the PFs can be rigorously defined for systems with semi-simple eigenvalues by extending their formulations through eigenvalue sensitivity and residue-based interpretations (see Methods for the details of the derivation process). This extension removes a long-standing restriction that has been imposed on the previous PF analyses and broadens their applicability to converter-dominated power systems. In such systems, the assumption of simple eigenvalues has often been implicitly adopted yet rarely examined or explicitly verified (see Supplementary Information Section 3 for the application of the PF).

Nevertheless, even with this generalized definition, the PFs remain confined to fixed spaces and therefore cannot explain the reconfiguration of the mode–state interactions observed under small perturbations. Consequently, the mechanisms by which perturbations reorganize the interactions between oscillatory modes and controller states remain mathematically unresolved. In particular, although the PF-based tuning strategies may improve the stability of systems, they do not reveal how mode–state interactions are reconfigured under perturbations. Such unobserved reconfiguration of interactions can fundamentally reshape controller behavior and, ultimately, system stability.

To resolve this remaining theoretical gap, we introduce an interaction sensitivity framework formulated in perturbation space that characterizes the variations in PFs defined between fixed spaces under perturbations, as shown in Fig. 1. This framework explicitly describes how mode–state interactions are reconfigured in response to perturbations. Rather than evaluating PFs within fixed spaces at given operating points, our proposed framework tracks the sensitivity of interactions as the system matrix changes under perturbations. This formulation enables the quantification of how oscillatory modes and controller states jointly respond to perturbations and how their mutual influence is reconfigured under perturbation-driven evolution processes. By shifting the analysis perspective from a fixed space to the perturbation space, the framework reveals interaction pathways and causal relationships that are fundamentally inaccessible to the conventional PF analysis technique. As a result, it provides a unified and physically interpretable description of stability evolution trends under perturbations, laying the foundation for systematic controller designs and interaction analyses involving complex systems.

## Participation factor sensitivity

Motivated by the limitations of the conventional PF analysis methods discussed above, we derive the mathematical meaning of PF variation. The PF variation for the $k$-th state and the $i$-th mode is expressed as

$$\mathbf{PF}_{k,i}(\varepsilon)-\mathbf{PF}_{k,i}=\mathbf{U}_{k,i}(\varepsilon)\cdot\mathbf{W}_{i,k}(\varepsilon)-\mathbf{U}_{k,i}\cdot\mathbf{W}_{i,k} \tag{1}$$

where $\mathbf{PF}_{k,i}(\varepsilon)$ is the PF of a perturbed system matrix $\mathbf{A}(\varepsilon) \in \mathbb{R}^{n\times n}$ (see Methods for details of the derivation). Additionally, $\mathbf{A}(\varepsilon) = \mathbf{A} + \varepsilon \cdot \mathbf{B} \in \mathbb{R}^{n\times n}$, where both $\mathbf{A}$ and $\mathbf{A}(\varepsilon)$ are assumed to be diagonalizable, and $\varepsilon \cdot \mathbf{B}$ is a perturbation matrix with a small parameter $\varepsilon$. $\mathbf{U}(\varepsilon)$, $\mathbf{W}(\varepsilon) \in \mathbb{C}^{n\times n}$ are matrices of the right and left eigenvectors derived from $\mathbf{A}(\varepsilon)$. Note that the right and left eigenvectors are biorthogonally normalized. Because $\mathbf{A}$ and $\mathbf{A}(\varepsilon)$ originate from differential–algebraic equations, the eigenvectors and, therefore, the PF variations are, in general, multivariate functions of the state and algebraic variables. Here, we reinterpret these multivariate functions in terms of $\varepsilon$ by expressing them as functions of $\varepsilon$. Under this formulation, the PF variation is characterized by its sensitivity to $\varepsilon$. The sensitivity of the PFs with respect to $\varepsilon$ then provides a quantitative measure of how mode–state interactions evolve under perturbations. By ignoring the higher-order terms with exponents greater than 1, we define the participation factor sensitivity (PFS) with the first-order approximation of the PF variation as

$$\mathbf{PF}_{k,i}(\varepsilon)-\mathbf{PF}_{k,i}\approx\varepsilon\cdot\left.\frac{d\mathbf{PF}_{k,i}}{d\varepsilon}\right|_{\varepsilon=0}\triangleq\mathbf{PFS}_{k,i} \tag{2}$$

Because the PF variation is subjected to $\varepsilon$, the event-dependent nature of the system can be fully understood, which has not been achieved in previous PF analyses conducted in fixed spaces. Therefore, as shown in Fig. 2, we newly propose the concept of the PFS in perturbation space to analyze the unknown phenomenon occurring between two fixed spaces. The evaluation of the PFS is based on analytic perturbation theory, which provides the perturbation series of the eigensolutions for both simple and semi-simple eigenvalues, which are expressed as the Taylor and Puiseux series, respectively[30-37] (see Supplementary Information Section 4). Hence, the PFSs for both eigenvalue cases can be calculated by using equation (2), while neglecting the higher-order terms.

As the PFs quantify the interactions between the modes and states, their sensitivity inherently contains information about how such interactions are reconfigured under perturbations. To reveal this interaction mechanism, we introduce two distinct yet complementary quantities—modal interaction sensitivity (MIS) and state interaction sensitivity (SIS)—which originate from the same PFS and provide physically interpretable interaction reconfiguration perspectives.

**Modal interaction sensitivity**

The PFS captures how mode–state interactions vary under perturbations. To further elucidate how these interactions are reorganized, we introduce the MIS, which quantifies the perturbation-driven coupling between two modes. Unlike the classic modal interaction indices (MIIs) reported in refs. 38,39, the MIS explicitly incorporates perturbation-driven sensitivity, enabling an event-dependent interpretation of modal interactions in perturbation space.

We first consider the case with simple eigenvalues, for which the interpretation of the MIS can be directly related to the first-order perturbation of eigenvectors. For a simple eigenvalue, we interpret the PFS as the first-order term in the multiplication of the perturbation series of the right and left eigenvectors (see Supplementary Information Section 4 for details regarding the perturbation series of the eigenvectors and Methods for a brief derivation of the PFS) as

$$\mathbf{PFS}_{k,i}=\sum_{\substack{j=1\\ j\neq i}}^{n}\left[\left(\frac{\varepsilon\cdot\mathbf{W}_i^{*}\cdot\mathbf{B}\cdot\mathbf{U}_j}{\lambda_i-\lambda_j}\right)\cdot\left(\mathbf{U}_{k,i}\cdot\mathbf{W}_{j,k}\right)+\left(\frac{\varepsilon\cdot\mathbf{W}_j^{*}\cdot\mathbf{B}\cdot\mathbf{U}_i}{\lambda_i-\lambda_j}\right)\cdot\left(\mathbf{U}_{k,j}\cdot\mathbf{W}_{i,k}\right)\right] \tag{3}$$

This formulation reveals that the interaction structure captured by the PFS extends from mode–state interactions to interaction sensitivities between pairs of modes. In equation (3), the PFS is expressed as the sum of a linear combination of several quantities. In particular, the paired quantities ($\mathbf{U}_{k,i}\cdot\mathbf{W}_{j,k}$) and ($\mathbf{U}_{k,j}\cdot\mathbf{W}_{i,k}$) form the classic MII, which is defined as $\mathbf{MII}_j(k,i)=\mathbf{U}_{k,i}\cdot\mathbf{W}_{j,k}+\mathbf{U}_{k,j}\cdot\mathbf{W}_{i,k}$. We interpret these paired terms as duals and modify them as $\mathbf{MII}_{j,i}^{(L)}(k,i)$ and $\mathbf{MII}_{j,i}^{(R)}(k,i)$, respectively, where the superscripts ($L$) and ($R$) indicate distinct modal interaction mechanisms. Then, the MIIs represent the interactions between the $i$-th and $j$-th modes when the $k$-th state is fixed. However, because the MIIs are computed by using unperturbed right and left eigenvectors defined in fixed space, they do not capture how sensitive the modal interactions are to a perturbation. To overcome this limitation, the PFS is formulated by using the modal sensitivity coefficients (MSCs), which are $\mathbf{MSC}_{j,i}^{(L)}(k,i)=\left(\frac{\varepsilon\cdot\mathbf{W}_i^{*}\cdot\mathbf{B}\cdot\mathbf{U}_j}{\lambda_i-\lambda_j}\right)$ and $\mathbf{MSC}_{j,i}^{(R)}(k,i)=\left(\frac{\varepsilon\cdot\mathbf{W}_j^{*}\cdot\mathbf{B}\cdot\mathbf{U}_i}{\lambda_i-\lambda_j}\right)$ as

$$\mathbf{PFS}_{k,i}=\sum_{\substack{j=1\\ j\neq i}}^{n}\left[\mathbf{MSC}_{j,i}^{(L)}(k,i)\cdot\mathbf{MII}_{j,i}^{(L)}(k,i)+\mathbf{MSC}_{j,i}^{(R)}(k,i)\cdot\mathbf{MII}_{j,i}^{(R)}(k,i)\right] \tag{4}$$

Note that $\varepsilon\cdot\mathbf{B}$ in the MSCs is coupled with the right and left eigenvectors. This enables the MSCs to act as event-dependent weighting factors. Furthermore, they are sensitive to the per-turbation if the two modes, $\lambda_i$ and $\lambda_j$, are close to each other. As a result, the event-dependent sensitivity of the modal interaction is measured in perturbation space, where the PFs evolve under perturbations. The MIS is then defined by combining the MIIs and the MSCs, yielding the relationship between the PFS and the MIS as

$$\mathbf{PFS}_{k,i} = \sum_{\substack{j=1 \\ j \neq i}}^{n} \mathbf{MIS}_j\left(k, i\right) \tag{5}$$

We next extend the definition of the MIS to systems with semi-simple eigenvalues, which commonly arise in converter-dominated power systems. The MIS for a semi-simple eigenvalue can be expressed as

$$\begin{aligned}\mathbf{PFS}_{k,i} &= \sum_{j \notin M_i} \sum_{h \in M_i} \left[ \left( \mathbf{L}_{h,i} \cdot \frac{\varepsilon \cdot \mathbf{W}_h^* \cdot \mathbf{B} \cdot \mathbf{U}_j}{\lambda_i - \lambda_j} \right) \cdot \left( \mathbf{U}_{k,i} \cdot \mathbf{W}_{j,k} \right) \right. \\ &\qquad \left. + \left( \mathbf{R}_{h,i} \cdot \frac{\varepsilon \cdot \mathbf{W}_j^* \cdot \mathbf{B} \cdot \mathbf{U}_h}{\lambda_i - \lambda_j} \right) \cdot \left( \mathbf{U}_{k,j} \cdot \mathbf{W}_{i,k} \right) \right] \\ &= \sum_{j \notin M_i} \sum_{h \in M_i} \left[ \mathbf{MSC}_{j,h}^{(L)}\left(k,i\right) \cdot \mathbf{MII}_{j,h}^{(L)}\left(k,i\right) + \mathbf{MSC}_{j,h}^{(R)}\left(k,i\right) \cdot \mathbf{MII}_{j,h}^{(R)}\left(k,i\right) \right] \\ &= \sum_{j \notin M_i} \mathbf{MIS}_j\left(k,i\right)\end{aligned} \tag{6}$$

where $\mathbf{L}_{h,\,i}$ and $\mathbf{R}_{h,\,i}$ are the solutions in the null space of linear equations related to the left and right eigenvectors, respectively, except for zero vectors (see Supplementary Information Section 4). Additionally, $M_i = \{h|\lambda_h = \lambda_i,\ h \in \{1, \ldots, \dim(\mathbf{A})\}\}$ is the index set of repeated modes. By comparing equations (3) and (6), the paired terms of the MIIs satisfy $\mathbf{MII}_{j,\,i}^{(L)}\ (k, i) = \mathbf{MII}_{j,\,h}^{(L)}\ (k, i) = \mathbf{U}_{k,\,i} \cdot \mathbf{W}_{j,\,k}$ and $\mathbf{MII}_{j,\,i}^{(R)}\ (k, i) = \mathbf{MII}_{j,\,h}^{(R)}\ (k, i) = \mathbf{U}_{k,\,j} \cdot \mathbf{W}_{i,\,k}$, thereby preserving the modal interaction structure. Although $\varepsilon \cdot \mathbf{B}$ in the MSCs is coupled with the $h$-th mode rather than the $i$-th mode, the equality $\lambda_i = \lambda_h$ ensures that the effects of semi-simple eigenvalues are consistently captured. The contribution of the modal interactions to the PFS mainly depends on $\mathbf{L}_{h,\,i}$ and $\mathbf{R}_{h,\,i}$, which regulate the MSCs, and therefore, the MIS. Consequently, $\mathbf{MIS}_j\ (k, i)$ in equation (6) is computed by considering the influences of all repeated modes, whose indices are in $M_i$.

There are other important features in equations (3)–(6). The MSCs correspond to the first-order perturbations of eigenvectors in the power series (see Supplementary Information Section 4). Moreover, because the denominators of the MSCs must be non-zero (i.e., $\lambda_i \neq \lambda_j$), the MIS is defined only for distinct eigenvalue pairs. For consistency, the corresponding MII and MIS terms for $j \in M_i$ are excluded from the summations and assigned a value of zero only in the numerical implementation and visualizations. To yield concrete insights, the magnitudes of $\mathbf{MIS}_j\ (k, i)$ indicate how strongly the connection between the center and surrounding spheres for the $i$-th and $j$-th modes in the $k$-th state is established, as shown on the top right of Fig. 2. Depending on $\varepsilon \cdot \mathbf{B}$, the more connection branches there are, the more modal interactions act sensitively in perturbation space. In summary, the MSCs weight the contributions of the MIIs. Thus, the MIIs regulated via the MSCs, which correspond to the MISs, constitute the PFS for a given controller.

**State interaction sensitivity**

Similar to the MIS, the interactions between two states can be analyzed by interpreting the PFS physically. The MSC, MII, and MIS are calculated by using equations (3)–(6) when two modes interact through a given state. Moreover, the summation indices $j$ and $h$ represent modes $\lambda_j$ and $\lambda_h$, respectively, while ensuring that the sum of the MIS terms is equal to the PFS. For state interactions, we reconstruct the mathematical and physical approaches used to analyze the MIS by modifying its summation indices. In other words, a mode must be specified, and the summation indices must be assigned to states. Then, for a simple eigenvalue, we rewrite the PFS to characterize the SIS as

$$\mathbf{PFS}_{k,i} = \sum_{l=1}^{n} \left[ \left( \sum_{\substack{j=1 \\ j \neq i}}^{n} \frac{\varepsilon \cdot \tilde{\mathbf{U}}_{l,j} \cdot \mathbf{W}_{j,k}}{\lambda_i - \lambda_j} \right) \cdot \left( \mathbf{U}_{k,i} \cdot \mathbf{W}_{i,l} \right) + \left( \sum_{\substack{j=1 \\ j \neq i}}^{n} \frac{\varepsilon \cdot \mathbf{U}_{k,j} \cdot \tilde{\mathbf{W}}_{j,l}}{\lambda_i - \lambda_j} \right) \cdot \left( \mathbf{U}_{l,i} \cdot \mathbf{W}_{i,k} \right) \right] \qquad (7)$$

where $\widetilde{\mathbf{U}} = \mathbf{B} \cdot \mathbf{U}$ and $\widetilde{\mathbf{W}}^* = \mathbf{W}^* \cdot \mathbf{B}$. They are used to simplify the notation. We analyze equation (7) by using the logical flow introduced in the previous section. The paired terms $(\mathbf{U}_{k,i} \cdot \mathbf{W}_{i,l})$ and $(\mathbf{U}_{l,i} \cdot \mathbf{W}_{i,k})$ capture the interactions between the $k$-th and $l$-th states when the $i$-th mode is fixed. They are defined as the generalized participation (GP) (ref. 10) and the parallel form of the GP (ref. 15), respectively. For consistency, we newly define them as state interaction indices (SIIs), which are $\mathbf{SII}_{l,i}^{(L)}\ (k, i)$ and $\mathbf{SII}_{l,i}^{(R)}\ (k, i)$, implying duality. Because the SIIs are obtained by multiplying the elements of the eigenvectors, they exist in fixed space. For this reason, SIIs cannot be solely used to interpret the sensitivity of state interactions. Thus, the sensitivity analysis performed in perturbation space requires the coefficients that weight the SIIs. We define them as state sensitivity coefficients (SSCs), which are $\mathbf{SSC}_{l,i}^{(L)}(k,i) = \left(\sum_{j\neq i} \frac{\varepsilon \cdot \tilde{\mathbf{U}}_{l,j} \cdot \mathbf{W}_{j,k}}{\lambda_i - \lambda_j}\right)$ and $\mathbf{SSC}_{l,i}^{(R)}(k,i) = \left(\sum_{j\neq i} \frac{\varepsilon \cdot \mathbf{U}_{k,j} \cdot \tilde{\mathbf{W}}_{j,l}}{\lambda_i - \lambda_j}\right)$. Then, we can rewrite equation (7) as

$$\mathbf{PFS}_{k,i} = \sum_{l=1}^{n}\left[\mathbf{SSC}_{l,i}^{(L)}(k,i)\cdot\mathbf{SII}_{l,i}^{(L)}(k,i)+\mathbf{SSC}_{l,i}^{(R)}(k,i)\cdot\mathbf{SII}_{l,i}^{(R)}(k,i)\right] \tag{8}$$

Unlike the MSCs, $\varepsilon \cdot \mathbf{B}$ is not coupled with the right and left eigenvectors. Instead, the eigenvectors are weighted by $\varepsilon \cdot \mathbf{B}$ (see the definitions of $\tilde{\mathbf{U}}$ and $\tilde{\mathbf{W}}^*$). Furthermore, the SSCs are evaluated by summing the state interactions through $\varepsilon \cdot \mathbf{B}$, while all modes are considered via the summation index. For example, $\tilde{\mathbf{U}}_{l,j} \cdot \mathbf{W}_{j,k} = \mathbf{B}_{l,1} \cdot \mathbf{U}_{1,j} \cdot \mathbf{W}_{j,k} + \cdots + \mathbf{B}_{l,n} \cdot \mathbf{U}_{n,j} \cdot \mathbf{W}_{j,k}$ holds. Here, we observe the influence of state sensitivity by summing all state interactions involving the $k$-th state via $\varepsilon \cdot \mathbf{B}$. Additionally, modal effects can be analyzed within the SSCs due to the summation operation because the summation index is related to the mode. Therefore, the perturbation space enables direct quantification of the event-dependent sensitivity of the state interaction as the PF evolves. This sensitivity is defined as the SIS, and the relationship between the PFS and the SIS is formulated as

$$\mathbf{PFS}_{k,i} = \sum_{l=1}^{n}\mathbf{SIS}_l(k,i) \tag{9}$$

Similarly, the SIS for a semi-simple eigenvalue can be expressed as

$$\begin{aligned}\mathbf{PFS}_{k,i} &= \sum_{l=1}^{n}\sum_{h\in M_i}\left[\left(\sum_{j\notin M_i}\mathbf{L}_{h,i}\cdot\frac{\varepsilon\cdot\tilde{\mathbf{U}}_{l,j}\cdot\mathbf{W}_{j,k}}{\lambda_i-\lambda_j}\right)\cdot\left(\mathbf{U}_{k,i}\cdot\mathbf{W}_{h,l}\right)\right.\\ &\qquad\left.+\left(\sum_{j\notin M_i}\mathbf{R}_{h,i}\cdot\frac{\varepsilon\cdot\mathbf{U}_{k,j}\cdot\tilde{\mathbf{W}}_{j,l}}{\lambda_i-\lambda_j}\right)\cdot\left(\mathbf{U}_{l,h}\cdot\mathbf{W}_{i,k}\right)\right]\\ &= \sum_{l=1}^{n}\sum_{h\in M_i}\left[\mathbf{SSC}_{l,h}^{(L)}(k,i)\cdot\mathbf{SII}_{l,h}^{(L)}(k,i)+\mathbf{SSC}_{l,h}^{(R)}(k,i)\cdot\mathbf{SII}_{l,h}^{(R)}(k,i)\right]\\ &= \sum_{l=1}^{n}\mathbf{SIS}_l(k,i)\end{aligned} \tag{10}$$

When evaluating the SIIs, the mode must be fixed according to the definition. In equation (10), although $i$ and $h$ denote different indexed modes, they correspond to the same repeated eigenvalue, satisfying $\lambda_i = \lambda_h$, which ensures that the state interactions ($\mathbf{U}_{k,i} \cdot \mathbf{W}_{h,l}$) and ($\mathbf{U}_{l,h} \cdot \mathbf{W}_{i,k}$) are well defined. In addition, the SSCs have the linear combination coefficients $\mathbf{L}_{h,i}$ and $\mathbf{R}_{h,i}$, similar to what is defined in the MIS, while indicating whether the effects of the semi-simple eigenvalues are dominant. Interestingly, because there is no restriction on the summation index $l$, the self-interaction sensitivity of a state can be derived as $\mathbf{SIS}_k$ ($k$, $i$). As a result, the self-interaction of a state can also be defined in the SII.

In particular, if the SIIs in equations (8) and (10) satisfy $l = k$ and $h = i$, they become $\mathbf{U}_{k,\,i} \cdot \mathbf{W}_{i,\,k}$, which is the same as $\mathbf{PF}_{k,\,i}$. For a comprehensive understanding of the SIS, it is important to consider the magnitudes of the SIS terms. These magnitudes represent how strongly the connection between the center and surrounding spheres for the $k$-th and $l$-th states in the $i$-th mode is established. This is illustrated on the bottom right of Fig. 2, where the self-interaction sensitivity for the $k$-th state, $\mathbf{SIS}_k$ ($k$, $i$), is also shown. From a mathematical and physical perspective, our results reveal an interpretive duality between the MIS and the SIS. Both quantities are derived from the PFS, so, as mentioned, their sums are equal to the PFS. That is, with the appropriate exclusion of repeated-mode indices in the MIS summation, $\sum_{j \notin M_i} \mathbf{MIS}_j\ (k, i) = \sum_l \mathbf{SIS}_l\ (k, i) = \mathbf{PFS}_{k,\,i}$. By changing the eigenvectors and summation indices, the PFS in perturbation space can be represented as the MIS and SIS, interpreting different interaction mechanisms. In contrast, the MII and SII, constituting the MIS and SIS terms, respectively, are defined only in fixed space. Understanding how the systems evolve across two fixed spaces provides insight into the dynamics and tendencies of internal controller actions and their effect on overall stability. This unique interaction sensitivity framework for the PFS, MIS, and SIS provides a basis for such analysis.

**Interaction sensitivity analysis in perturbation space**

To assess the three sensitivity measures, we construct a converter-dominated power system with wind turbines, as shown in Fig. 3a, b. Many studies have reported that a poorly tuned phase-locked loop (PLL) causes instability[40-43]. Thus, for a small perturbation (to satisfy the linearization requirement), we vary the value of the proportional gain of the synchronous reference frame PLL (SRF-PLL) controller, which is $k_{p,pll}$, from 5 to 50, while keeping the value of the integral gain, $k_{i,pll}$, at 5000. The other variables are held constant (see Extended Data Fig. 1 for a detailed control block diagram of the SRF-PLL controller). As shown in Fig. 3c, the real parts of all eigenvalues of $\mathbf{A}$ and $\mathbf{A}$ ($\varepsilon$) are smaller than or equal to zero. The algebraic and geometric multiplicities of each eigenvalue are the same, which results in a globally stable system according to the conventional stability theory[44]. For a sudden load change, we also confirm that increasing $k_{p,pll}$ makes the system more stable in the time and frequency domains by mitigating the oscillatory mode (see Extended Data Fig. 2). The magnitudes of the PFS are calculated by using equation (2), and the results are plotted in Fig. 3d (see Extended Data Fig. 3 for the PFs before and after the perturbation). The green elements indicate that the PFs have varied in the transition between the two fixed spaces, whereas most PFS elements are nearly zero (shown in white). In particular, the degree of green intensity represents the extent of the variation in PFs. This demonstrates that the perturbation selectively reconfigures the interactions associated with specific oscillations and controllers rather than inducing global variations across the system. Although the PFS is formulated by the first-order approximation of the variation in the PFs, the absolute error between the exact PF variation and the PFS is sufficiently small for the considered perturbation. (see Extended Data Fig. 4). Notably, the PF and eigenvalue analyses are powerful methods for assessing the stability issues encountered before and after the perturbation. However, these methods cannot reveal whether the interaction structure of the system is considerably reconfigured, which is a feature that cannot be inferred solely from the PFs and eigenvalues.

When analyzing the MIS and SIS, we focus on the dominant mode $\lambda_{39}$ with the smallest damping ratio, and $x_{18}$, which is one of the SRF-PLL states (see Extended Data Table 1 for detailed nomenclature about the states). The states of the SRF-PLL highly affect the system stability because their magnitudes of the PFs in $\lambda_{39}$ are considerably larger than those of other PFs (i.e., $|\mathbf{PF}_{19,39}|$ ($\approx 0.4868$) > $|\mathbf{PF}_{18,39}|$ ($\approx 0.4862$) ≫ $|\mathbf{PF}_{20,39}|$ ($\approx 0.0138$) > ...) (see Extended Data Fig. 3). Under this condition, the MIIs in Fig. 4a, b indicate that various modal interactions occur in fixed space. However, because these values exist before the perturbation, they cannot explain the event-dependent characteristics of the system. On the other hand, the MSCs in Fig. 4c, d provide information about the perturbation by capturing the evolution of the modal interactions. As a result, the MIS in Fig. 4e represents the contribution of the variation of the modal interaction in perturbation space. It also shows that the most affected mode in $\mathbf{PFS}_{18,39}$ is $\lambda_{38}$ with a magnitude of approximately 0.1587, which is the complex conjugate pair of $\lambda_{39}$. Note that a perturbation generally induces changes in PFs. Thus, if we can identify one of the PF elements that should be analyzed, we can determine which modal interactions in the specific controller vary most sensitively by using the MIS. This enables the direct identification of which modal interactions are most sensitive to a given perturbation, which is a capability that is fundamentally unavailable in classic PF analyses. Notably, the MIS inherits the complex conjugate symmetry of the underlying linear system. For a complex conjugate eigenvalue pair $\alpha \pm j\beta$, with $\beta > 0$, the corresponding modal terms combine in a time-domain response to form a single oscillatory component with exponential rate $\alpha$ and angular frequency $\beta$, rather than two distinct oscillations. Consistent with this symmetry, the complex conjugate of the MIS formed by two modes with different oscillations and damping properties is the MIS formed by their conjugate modes (e.g., $\mathbf{MIS}_{36}\,(18, 39) = \overline{\mathbf{MIS}_{37}\,(18, 38)}$, $\lambda_{36} = \bar{\lambda}_{37}$, and $\lambda_{38} = \bar{\lambda}_{39}$) (see Extended Data Fig. 5 for a detailed illustration). Together, the MII, MSC, and MIS described above offer a new perspective for both interpreting the PFS and uncovering more profound modal interactions.

The SIIs in Fig. 5a, b indicate that numerous state interactions are observed in fixed space. Also, the impact of the perturbation on the state interactions can be quantified on the basis of the SSCs in Fig. 5c, d. By combining those two characteristics, we can identify previously undetectable controller dynamics with respect to the PF in perturbation space, as shown in Fig. 5e. The most influential state interacting with $x_{18}$ in $\lambda_{39}$ is $x_{19}$, where both states are related to the SRF-PLL controller. This observation can be attributed to the intrinsic dynamical dependence between $x_{18}$ and $x_{19}$ imposed by the SRF-PLL control structure (see Extended Data Fig. 1). Additionally, the SISs of $x_{20}$ and $x_{21}$, which are the states associated with the low-pass filter in Fig. 3b, have the next largest values. This is because the wind turbine is modeled as a voltage source whose input signals are related to the output signals of the SRF-PLL controller. This means that, although the controllers are separate from each other, the characteristics of the wind turbine are reflected in the SIS. Moreover, we show the agreement with the basis of a linear system mentioned earlier by comparing the SIS results for a complex conjugate pair of modes (e.g., $\mathbf{SIS}_{19}\,(18, 39) = \overline{\mathbf{SIS}_{19}\,(18, 38)}$ and $\lambda_{38} = \bar{\lambda}_{39}$) (see Extended Data Fig. 6 for a detailed illustration). Therefore, our proposed analysis reveals how the perturbation influences the connectivity among the controllers in an oscillatory mode.

## Discussion

The high penetration of converter-interfaced RESs fundamentally alters how stability must be understood in modern power systems. Recent large-scale disturbance events have highlighted the fact that stability challenges increasingly stem from perturbation-driven reconfigurations of the dynamic interactions among oscillatory modes and controller state vari-ables. Despite their long-standing utility, classic PF and eigenvalue analyses are intrinsically confined to fixed operating points and therefore unable to capture event-dependent interaction variations. This limitation motivates a conceptual perspective shift, expanding the stability analysis paradigm beyond interactions in a fixed space to a perturbation space where interaction reconfigurations can be systematically analyzed.

In this work, we introduce an interaction sensitivity framework that is formulated in perturbation space, providing analytic expressions and their physical properties. The essence of this framework is the PFS, which quantifies the PF variations induced by perturbations. Unlike PF analyses performed at steady-state operating points, the PFS captures how mode–state interactions are reconfigured between two fixed spaces under perturbations. This perspective shift enables the direct investigation of the event-dependent interaction variations that are structurally inaccessible to the existing tools.

A key aspect of the proposed framework is the removal of a restrictive assumption underlying previous PF analyses: the implicit requirement of simple eigenvalues. Converter-dominated power systems frequently exhibit semi-simple eigenvalues arising from controller designs, symmetry structures, and operational constraints. By rigorously extending the definition of PFs to systems with semi-simple eigenvalues using eigenvalue sensitivity and residue-based interpretations, we establish a mathematically consistent foundation for conducting interaction analyses of high-dimensional systems. While this generalized PF formulation remains confined to fixed space, it provides the mathematical underpinning for the proposed interaction sensitivity framework to operate without relying on the implicit assumption of simple eigenvalues.

The MIS and SIS provide distinct yet complementary descriptions of perturbation-driven interaction reconfigurations, both originating from the same PFS. From a modal interaction perspective, our framework shows how perturbations affect the coupling between oscillatory modes for a given controller state. From a state interaction perspective, it reveals how controller states reorganize their mutual influences in response to perturbations within a specific oscillatory mode. These dual interpretations demonstrate that perturbation-driven stability evolution trends can be analyzed via different interaction pathways. Importantly, both perspectives maintain mathematical consistency within a unified framework in perturbation space.

The case study performed in this work illustrates the practical implications of this conceptual shift. In the converter-dominated power system integrated with wind turbines, an eigenvalue analysis indicates that the system is globally stable before and after a perturbation occurs. In contrast, the interaction sensitivity framework reveals a clear reconfiguration of the mode–state interactions. Specifically, the PFS indicates the fact that several modes and states represent strong interaction sensitivities in perturbation space. This highlights that the perturbation selectively reshapes interaction structures rather than inducing variations across the entire system.

To further illustrate the applicability of the proposed framework, we consider two additional examples involving a spring–mass–damper system and a Conseil International des Grands Réseaux Electriques (CIGRE) medium-voltage microgrid benchmark system[45], as presented in Supplementary Information Sections 5 and 6, respectively. The first system is a fundamental model for studying mechanical vibration dynamics[46]. The second example, the microgrid benchmark system, is representative of systems used to analyze the effects of the high penetration of distributed energy resources (DERs), such as wind power plants (WPPs), solar photovoltaic plants, and energy storage systems[47]. In such a system, the stability challenges motivate the need for analytical tools capable of interpreting interaction mechanisms beyond conventional approaches. In both cases, the interaction sensitivities enable the identification of the modes and states that are affected by perturbations, providing engineering insights for targeted controller design and analysis.

Taken together, the results presented in this work establish interaction sensitivity in perturbation space as a unifying analytical and physical perspective for understanding stability beyond steady-state operating points. By characterizing how mode–state interactions are reconfigured under perturbations, the proposed framework connects conventional PF and eigenvalue analyses with the event-dependent nature of the system dynamics. Notably, this perspective makes it possible to systematically identify interaction mechanisms obscured in fixed space without relying on the implicit assumption of simple eigenvalues. As demonstrated, this framework facilitates an integrated understanding of perturbation-driven interaction reconfiguration, with broad implications for complex mechanical and electrical systems.

## Methods

### Derivation of the participation factor for simple eigenvalues

Consider a system matrix $\mathbf{A} \in \mathbb{R}^{n\times n}$ with non-repeated eigenvalues. The participation factor (PF) for the $k$-th state in the $i$-th mode (or, equivalently, the $i$-th mode in the $k$-th state) is calculated by multiplying the elements of linearly independent right and left eigenvectors as

$$\mathbf{PF}_{k,i} = \mathbf{U}_{k,i} \cdot \mathbf{W}_{i,k} \tag{11}$$

where $\mathbf{U}, \mathbf{W} \in \mathbb{C}^{n\times n}$ are matrices of the right and left eigenvectors, which can be described as a linear eigenvalue problem, $\mathbf{A} \cdot \mathbf{U} = \mathbf{U} \cdot \mathbf{\Lambda}$ and $\mathbf{W}^* \cdot \mathbf{A} = \mathbf{\Lambda} \cdot \mathbf{W}^*$ (i.e., $\mathbf{A}$ is diagonalizable). The right and left eigenvectors are biorthogonally normalized such that $\mathbf{W}^* \cdot \mathbf{U} = \mathbf{I}_n$. Additionally, $\mathbf{\Lambda} = \text{diag}\ (\lambda_1, \ldots, \lambda_n) \in \mathbb{C}^{n\times n}$ is a matrix of eigenvalues with diagonal elements, and the superscript * indicates the conjugate transpose operation. The subscript indices such as $k$ and $i$ represent the matrix entries.

### Derivation of the participation factor for semi-simple eigenvalues

We present eigenvalue sensitivity and residue-based formulations of PFs for semi-simple eigenvalues. Let $\mathbf{A}$ be diagonalizable and let its right and left eigenvectors be biorthogonally normalized. Let $\mu_1, \ldots, \mu_q$, with $q \leq n$, denote the distinct eigenvalues of $\mathbf{A}$, and define the corresponding index set as $M_\alpha = \{i|\lambda_i = \mu_\alpha\}$, with $m_\alpha = |M_\alpha|$. Any repeated eigenvalue of $\mathbf{A}$ is therefore a semi-simple eigenvalue and admits a complete set of linearly independent eigenvectors spanning the corresponding eigenspace. Once a biorthogonal basis is selected within each repeated eigenspace, the PF of each indexed mode is computed as $\mathbf{PF}_{k,i} = \mathbf{U}_{k,i} \cdot \mathbf{W}_{i,k}$. These individual PFs are defined with respect to the selected basis, whereas their sum over $M_\alpha$ is basis invariant and equals the $k$-th diagonal element of the eigenprojection associated with $\mu_\alpha$.

**Eigenvalue sensitivity.** For a repeated semi-simple eigenvalue $\mu_\alpha$, let the previously defined index set be ordered as $M_\alpha = \{i_1, \ldots, i_{m_\alpha}\}$. The corresponding right and left eigenvectors are collected as $\mathbf{U}_\alpha = [\mathbf{U}_{i_1}, \ldots, \mathbf{U}_{i_{m_\alpha}}]$ and $\mathbf{W}_\alpha^* = [\mathbf{W}_{i_1}, \ldots, \mathbf{W}_{i_{m_\alpha}}]^*$, where $\mathbf{W}_\alpha^* \cdot \mathbf{U}_\alpha = \mathbf{I}_{m_\alpha}$. To examine the sensitivity to the $k$-th diagonal element of $\mathbf{A}$, let $\mathbf{A}\ (\tau) = \mathbf{A} + \tau \cdot \mathbf{E}_{k,k}$, where $\tau$ is a scalar perturbation parameter, and $\mathbf{E}_{k,k}$ is the matrix whose element in the $k$-th row and $k$-th column is 1 and other elements are 0. Under this perturbation, the repeated eigenvalue $\mu_\alpha$ separates into $m_\alpha$ eigenvalue branches (see Supplementary Information Section 4). The first-order coefficients of these branches are the eigenvalues of the reduced eigenvalue sensitivity matrix $\mathbf{H}_{\alpha,k} = \mathbf{W}_\alpha^* \cdot \mathbf{E}_{k,k} \cdot \mathbf{U}_\alpha$. For the selected biorthogonal basis, the $r$-th diagonal element of $\mathbf{H}_{\alpha,k}$ can be expressed as

$$\left[\mathbf{H}_{\alpha,k}\right]_{r,r} = \mathbf{W}_{i_r}^* \cdot \mathbf{E}_{k,k} \cdot \mathbf{U}_{i_r} = \mathbf{U}_{k,i_r} \cdot \mathbf{W}_{i_r,k} = \mathbf{PF}_{k,i_r} \tag{12}$$

Therefore, the sensitivity of the sum of the eigenvalue branches can be calculated as

$$\frac{d}{d\tau}\sum_{r=1}^{m_\alpha} \lambda_{\alpha,r}(\tau)\Bigg|_{\tau=0} = \mathrm{tr}\left(\mathbf{H}_{\alpha,k}\right) = \sum_{i\in M_\alpha} \mathbf{PF}_{k,i} \tag{13}$$

Thus, each $\mathbf{PF}_{k,i}$ represents a basis-resolved diagonal contribution to the eigenvalue sensitivity, whereas their sum over the repeated eigenspace is basis invariant. Indeed, changing the biorthogonal basis within the repeated eigenspace transforms $\mathbf{H}_{\alpha,k}$ by a similarity transformation and therefore does not change its trace.

**Residue.** A residue-based interpretation can be obtained from a single-input single-output state-space system. Let $\mathbf{J} \in \mathbb{R}^{n\times 1}$ and $\mathbf{K} \in \mathbb{R}^{1\times n}$ be the input and output vectors, respectively. By using the eigendecomposition of $\mathbf{A}$, the state-space equation and its transfer function can be expressed as

$$\begin{aligned}
\dot{\mathbf{X}} &= \mathbf{A}\cdot\mathbf{X} + \mathbf{J}\cdot u, \quad y = \mathbf{K}\cdot\mathbf{X} \\
G(s) &= \mathbf{K}\cdot\mathbf{U}\cdot\left(s\cdot\mathbf{I}_n - \mathbf{\Lambda}\right)^{-1}\cdot\mathbf{W}^*\cdot\mathbf{J} \\
&= \mathbf{K}\cdot\left(\sum_{i=1}^{n}\frac{\mathbf{U}_i\cdot\mathbf{W}_i^*}{s-\lambda_i}\right)\cdot\mathbf{J} \\
&= \sum_{i=1}^{n}\frac{r_i^{modal}}{s-\lambda_i} = \sum_{\alpha=1}^{q}\frac{\gamma_{\mu_\alpha}}{s-\mu_\alpha} \\
r_i^{modal} &= \mathbf{K}\cdot\mathbf{U}_i\cdot\mathbf{W}_i^*\cdot\mathbf{J}, \quad \gamma_{\mu_\alpha} = \sum_{i\in M_\alpha} r_i^{modal}
\end{aligned} \tag{14}$$

where $r_i^{modal}$ denotes the indexed modal contribution associated with the $i$-th mode in the selected biorthogonal basis. For a simple eigenvalue, $r_i^{modal}$ is identical to the conventional residue. For a repeated semi-simple eigenvalue, the indexed modes in $M_\alpha$ share the same denominator $s - \mu_\alpha$. Therefore, $\gamma_{\mu_\alpha}$ is the corresponding coefficient in the partial fraction expansion. When $\gamma_{\mu_\alpha} \neq 0$, it becomes the scalar residue of $G(s)$ at the pole $s = \mu_\alpha$. By applying appropri-ate conditions for **J** and **K** such that their $k$-th elements are equal to 1, and all other elements are equal to 0, the indexed modal contribution can be written as $r_i^{modal} = \mathbf{U}_{k,i} \cdot \mathbf{W}_{i,k} = \mathbf{PF}_{k,i}$. Accordingly, $\gamma_{\mu_\alpha} = \sum_{i \in M_\alpha} \mathbf{PF}_{k,i}$ holds. Thus, $\mathbf{PF}_{k,i}$ represents the mode-resolved contribution to the residue in the transfer function with respect to the selected biorthogonal basis, whereas their sum gives the residue with the repeated eigenvalue.

The same interpretation can also be obtained directly from the resolvent of **A**, without specifying an external input and output pair. Note that **A** is the reduced system matrix obtained by linearizing the differential–algebraic equations (see Supplementary Information Section 1). In this case, the linear time-invariant system has the form of $\dot{\mathbf{X}} = \mathbf{A} \cdot \mathbf{X}$. By applying the eigendecomposition, **A** can be expressed as

$$\mathbf{A} = \sum_{i=1}^{n} \lambda_i \cdot \mathbf{U}_i \cdot \mathbf{W}_i^* \tag{15}$$

where equation (15) is referred to as the spectral representation of **A** (see Supplementary Information Section 4). The resolvent of **A** is written as

$$\left(s \cdot \mathbf{I}_n - \mathbf{A}\right)^{-1} = \frac{\operatorname{adj}\left(s \cdot \mathbf{I}_n - \mathbf{A}\right)}{\det\left(s \cdot \mathbf{I}_n - \mathbf{A}\right)} \tag{16}$$

where adj(·) and det(·) denote the adjugate matrix and determinant of a matrix, respectively. By using eigendecomposition, equation (16) can be rewritten as

$$\begin{aligned} \left(s \cdot \mathbf{I}_n - \mathbf{A}\right)^{-1} &= \mathbf{U} \cdot \left(s \cdot \mathbf{I}_n - \mathbf{\Lambda}\right)^{-1} \cdot \mathbf{W}^* \\ &= \sum_{i=1}^{n} \frac{\mathbf{U}_i \cdot \mathbf{W}_i^*}{s - \lambda_i} \\ &= \sum_{\alpha=1}^{q} \sum_{i \in M_\alpha} \frac{\mathbf{U}_i \cdot \mathbf{W}_i^*}{s - \mu_\alpha} \end{aligned} \tag{17}$$

The second equality in equation (17) resolves the residue expansion into indexed modal contributions with respect to the selected biorthogonal basis. When several indexed modes correspond to the same repeated semi-simple eigenvalue, their terms share the same denominator and collectively constitute the residue matrix associated with that pole. Accordingly, define

$$\mathbf{P}_{\mu_\alpha} = \sum_{i \in M_\alpha} \mathbf{U}_i \cdot \mathbf{W}_i^*, \ \left[\mathbf{P}_{\mu_\alpha}\right]_{k,k} = \sum_{i \in M_\alpha} \mathbf{PF}_{k,i} \tag{18}$$

Here, $\mathbf{P}_{\mu_\alpha}$ is the eigenprojection associated with $\mu_\alpha$ and the residue matrix of the resolvent at $s = \mu_\alpha$. The individual $\mathbf{PF}_{k,i}$ values are defined with respect to the selected biorthogonal basis, whereas $\mathbf{P}_{\mu_\alpha}$ and its diagonal elements are basis invariant. For a simple eigenvalue, $M_\alpha = \{i\}$ holds, so that $[\mathbf{P}_{\mu_\alpha}]_{k,k} = \mathbf{PF}_{k,i}$, representing the conventional residue-based PF formulation.

Combining equations (13) and (18) yields

$$\left.\frac{d}{d\tau}\sum_{r=1}^{m_\alpha}\lambda_{\alpha,r}(\tau)\right|_{\tau=0} = \mathrm{tr}\left(\mathbf{H}_{\alpha,k}\right) = \sum_{i\in M_\alpha}\mathbf{PF}_{k,i} = \left[\mathbf{P}_{\mu_\alpha}\right]_{k,k} \tag{19}$$

Therefore, the eigenvalue sensitivity and residue-based formulations yield the same basis-invariant PF sum while retaining a mode-resolved decomposition with respect to the selected biorthogonal basis.

**Analytic perturbation theory in the linear eigenvalue problem**

Let $\mathbf{A}(\varepsilon) \in \mathbb{R}^{n\times n}$ be a system matrix after perturbation with semi-simple eigenvalues (i.e., $\mathbf{A}(\varepsilon)$ is diagonalizable). Additionally, let $\mathbf{A}(\varepsilon) = \mathbf{A} + \varepsilon \cdot \mathbf{B}$, where $\varepsilon \cdot \mathbf{B}$ is a perturbation matrix with a small parameter $\varepsilon$. Then, a linear eigenvalue problem can be modeled as a perturbed system as

$$\begin{gathered}\mathbf{A}(\varepsilon)\cdot\mathbf{U}(\varepsilon) = \mathbf{U}(\varepsilon)\cdot\mathbf{\Lambda}(\varepsilon)\\ \mathbf{W}^*(\varepsilon)\cdot\mathbf{A}(\varepsilon) = \mathbf{\Lambda}(\varepsilon)\cdot\mathbf{W}^*(\varepsilon)\end{gathered} \tag{20}$$

where $\mathbf{U}(\varepsilon)$, $\mathbf{W}(\varepsilon) \in \mathbb{C}^{n\times n}$ are the matrices of the right and left eigenvectors derived from $\mathbf{A}(\varepsilon)$, respectively. For simple and semi-simple eigenvalues, the eigensolutions are derived from the Taylor series and Puiseux series, respectively, with continuously differentiable functions of $\varepsilon$ (see Supplementary Information Section 4). By using the perturbed eigenvectors, the PF of $\mathbf{A}(\varepsilon)$ for the $k$-th state in the $i$-th mode (or, equivalently, the $i$-th mode in the $k$-th state) can be computed as

$$\mathbf{PF}_{k,i}(\varepsilon) = \mathbf{U}_{k,i}(\varepsilon)\cdot\mathbf{W}_{i,k}(\varepsilon) \tag{21}$$

**Derivation of participation factor sensitivity for simple and semi-simple eigenvalues**

Suppose that $\lambda_i$ is an eigenvalue of the diagonalizable system matrix $\mathbf{A}$. Given that small perturbations occur in $\mathbf{A}$, the perturbation series of the right and left eigenvectors can be expanded as

$$\mathbf{U}_i(\varepsilon) \approx \mathbf{U}_i + \varepsilon\cdot\mathbf{U}_i' \tag{22}$$

$$\mathbf{W}_i^*(\varepsilon) \approx \mathbf{W}_i^* + \varepsilon \cdot \left(\mathbf{W}_i^*\right)' \tag{23}$$

where the higher-order terms are neglected (see Supplementary Information Section 4). Then, $\mathbf{PF}_{k,i}(\varepsilon)$ of $\mathbf{A}(\varepsilon)$ in equation (21) is computed by using equations (22) and (23) as

$$\begin{aligned}\mathbf{PF}_{k,i}(\varepsilon) &\approx \mathbf{U}_{k,i} \cdot \mathbf{W}_{i,k} + \varepsilon \cdot \left(\mathbf{U}_{k,i} \cdot \left(\mathbf{W}_{i,k}\right)' + \left(\mathbf{U}_{k,i}\right)' \cdot \mathbf{W}_{i,k}\right) \\ &= \mathbf{PF}_{k,i} + \varepsilon \cdot \left(\mathbf{U}_{k,i} \cdot \left(\mathbf{W}_{i,k}\right)' + \left(\mathbf{U}_{k,i}\right)' \cdot \mathbf{W}_{i,k}\right)\end{aligned} \tag{24}$$

By comparing equations (2) and (24), we obtain the PFS as

$$\mathbf{PFS}_{k,i} = \varepsilon \cdot \left(\mathbf{U}_{k,i} \cdot \left(\mathbf{W}_{i,k}\right)' + \left(\mathbf{U}_{k,i}\right)' \cdot \mathbf{W}_{i,k}\right) \tag{25}$$

Therefore, under the perturbation-consistent branch convention described in Supplementary Information Section 4, the expressions for the PFS in equations (3)–(6) are valid for both simple and semi-simple eigenvalues. Also, rearranging the terms in equations (3)–(6) with respect to the state variables yields equations (7)–(10).

**Modeling of the converter-dominated power system**

The converter-dominated power system in Fig. 3a, b is modeled by using DIg-SILENT PowerFactory® software. We utilize the Western Electricity Coordinating Council (WECC) generic model[48] developed for the WPP, which consists of 100 type-3 wind turbines, each rated at 2 MW. The real power of this equivalent aggregated model is modulated by the shaft speed, with the output of 150 MW. Additionally, the system is assumed to operate as a constant reactive power control method at 0 Mvar. Also, we add the LCL filters to the WPP to reduce the harmonics. Thereafter, the WPP is connected to an external power system, which is simply modeled as a single synchronous generator (SG) in the slack bus (i.e., the magnitude and phase angle of the voltage are specified). This SG includes excitation and turbine–governor systems, which are modeled as the IEEET1[49] and IEEEG1[50], respectively, with default parameters. The three-phase load is modeled as the constant power of 500 MW. Note that the steady-state operating condition is important because the purpose of linearization is to observe the system behavior at the equilibrium point. For this reason, when $\mathbf{A}$ and $\mathbf{A}(\varepsilon)$ are obtained before and after the perturbation, we must confirm whether the systems are stable.

**Data acquisition and explanation.** The built-in function of DIgSILENT PowerFactory® software was used to calculate the system matrices under steady-state operating conditions in the time-domain simulation. In MATLAB® software, we loaded the matrix information and used the eig.m function to compute the eigensolutions. All calculations for the PF, PFS, MIS, and SIS were also conducted in MATLAB® software.

**Data availability**

The data that support the findings of this study are available from the corresponding author upon reasonable request.

**Code availability**

The codes utilized during this study are available from the corresponding author upon reasonable request.

**Acknowledgements** This work was supported by the National Research Foundation (NRF) under Grants No. RS-2020-NR049406 and No. RS-2023-00218377 funded by the Ministry of Science and ICT (MSIT), South Korea.

**Author contributions** J.K. developed the concept of this study and performed the numerical simulations with help from J.-W.P. All authors prepared the manuscript. J.-W.P. supervised the project.

**Competing interests** The authors declare no competing interests.

**Additional information**

**Supplementary information** is available for this paper.

**Correspondence and requests for materials** should be addressed to Jung-Wook Park.

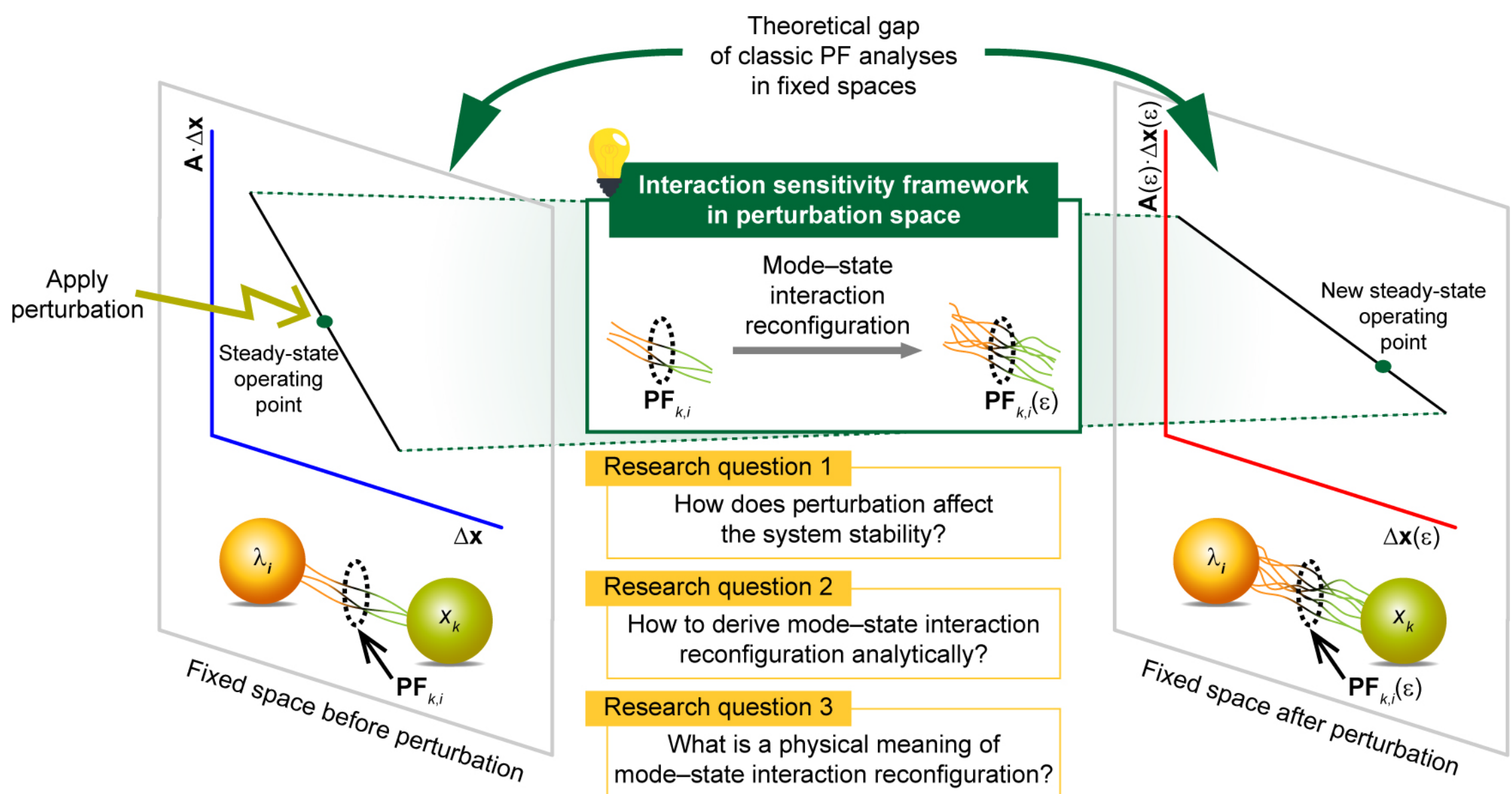

**Fig. 1. Overview of the interaction sensitivity framework in perturbation space.** When a perturbation occurs in a system operating in a steady-state, the system transitions to a new steady-state operating point. Therefore, the classic PF analyses are performed in fixed spaces before and after a perturbation occurs. In these spaces, the mode–state interactions, which are represented by the PFs, are evaluated by using the eigensolutions of $\mathbf{A}$ and $\mathbf{A}\,(\varepsilon)$. However, such analyses implemented in fixed spaces cannot capture the event-dependent evolution trends of interactions under perturbations. In contrast, the proposed interaction sensitivity framework introduces a perturbation space that characterizes event-dependent interaction reconfiguration between two fixed spaces. This perspective enables the analysis of stability evolution under perturbations and establishes a foundation for conducting physically interpretable interaction analyses beyond the conventional approaches.

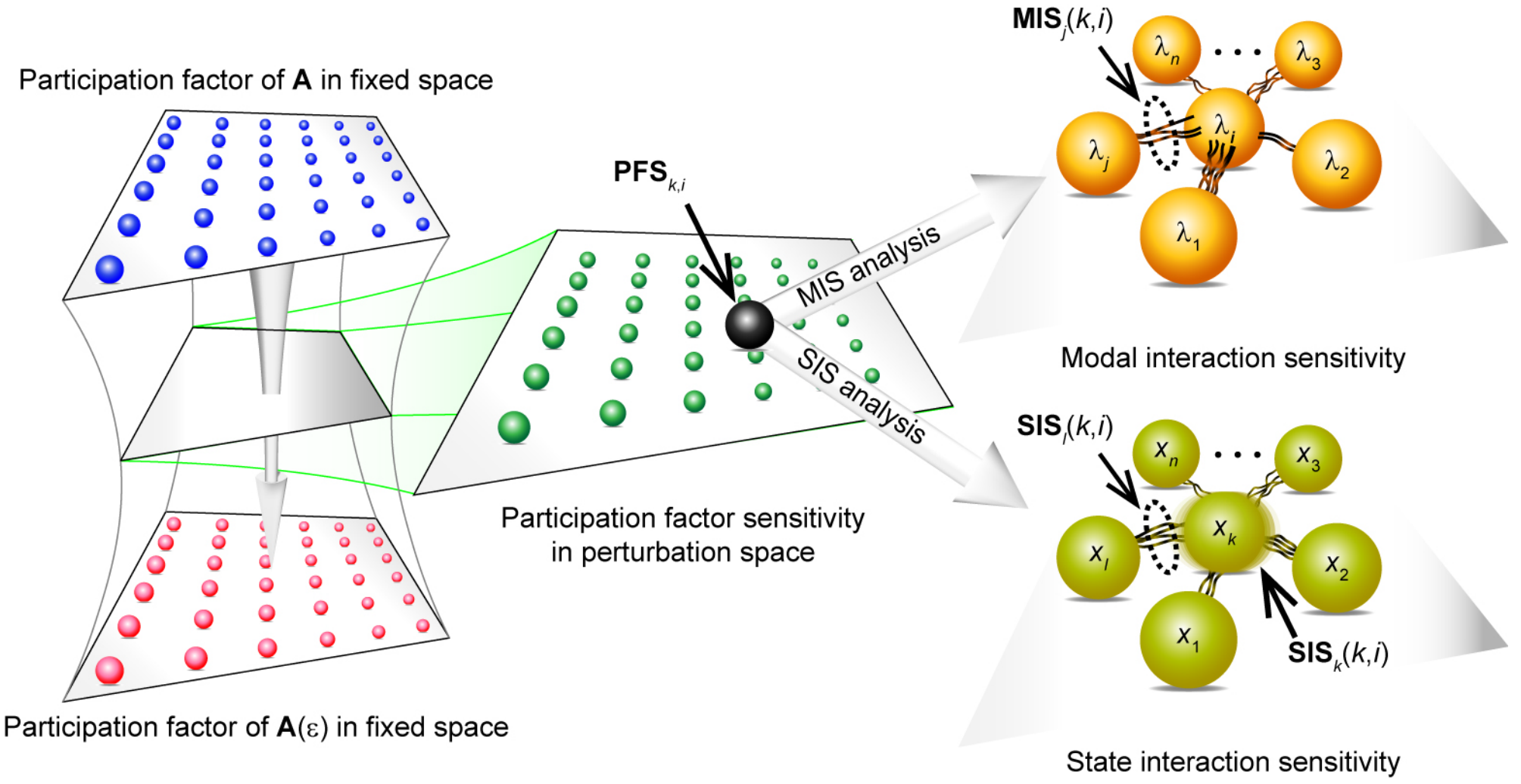

**Fig. 2. Schematic illustration of our proposed framework.** Two PFs can be generally obtained on the basis of right and left eigenvectors in fixed spaces before and after a perturbation occurs. On the other hand, the PFS is newly derived with right and left eigenvector perturbations in the transition between two fixed spaces. This enables MIS and SIS analyses to be performed. In other words, the MIS and SIS provide information on the strength of the physical connections in perturbation space. Unlike the MIS, the SIS additionally captures the self-interaction sensitivity of a state, which is denoted as $\mathbf{SIS}_k$ ($k$, $i$).

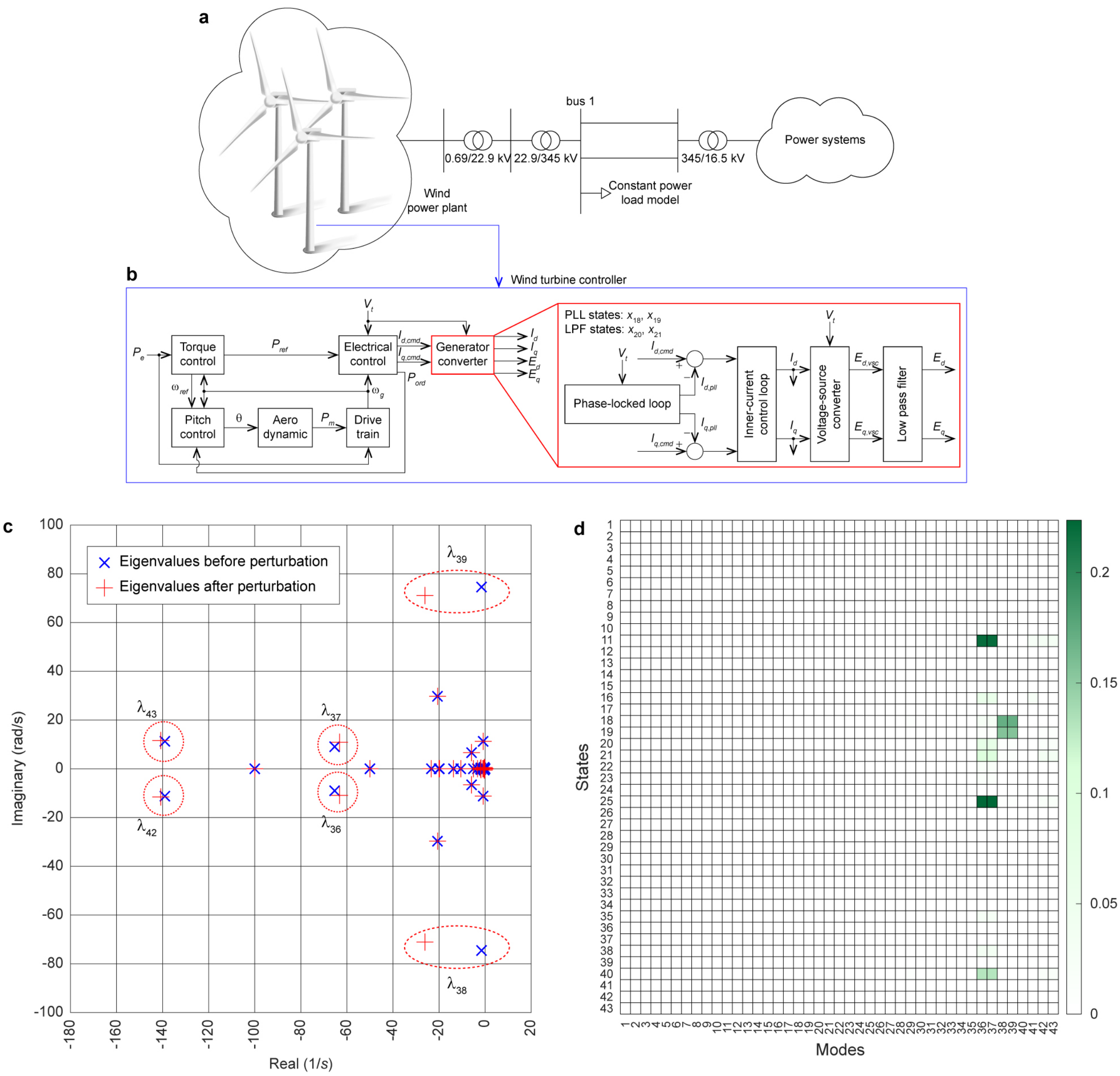

**Fig. 3. Example: Converter-dominated power system.** **a**, Configuration of the converter-dominated power system. All wind turbines in the wind power plant are assumed to operate on the basis of the real power modulated by the shaft speed, and the reactive power is constant. **b**, Control block diagram of each wind turbine system. The red box represents the interface between the converters and the electrical grid. Here, we change the value of the proportional gain of the SRF-PLL controller, $k_{p,pll}$, from 5 to 50, while keeping the value of the integral gain, $k_{i,pll}$, at 5000; all other variables remain fixed. **c**, Eigenvalues of **A** and **A** ($\varepsilon$) in the two fixed spaces, which are plotted with blue crosses and red plus markers, respectively. The variations in $\lambda_{36}$, $\lambda_{37}$, $\lambda_{38}$, $\lambda_{39}$, $\lambda_{42}$, and $\lambda_{43}$, plotted with dotted red circles, are larger than those in the other eigenvalues. **d**, Heatmap of the magnitudes of the PFSs corresponding to this small perturbation. Non-zero PFS elements indicate the selective reconfiguration of mode–state interactions, whereas most elements are nearly zero.

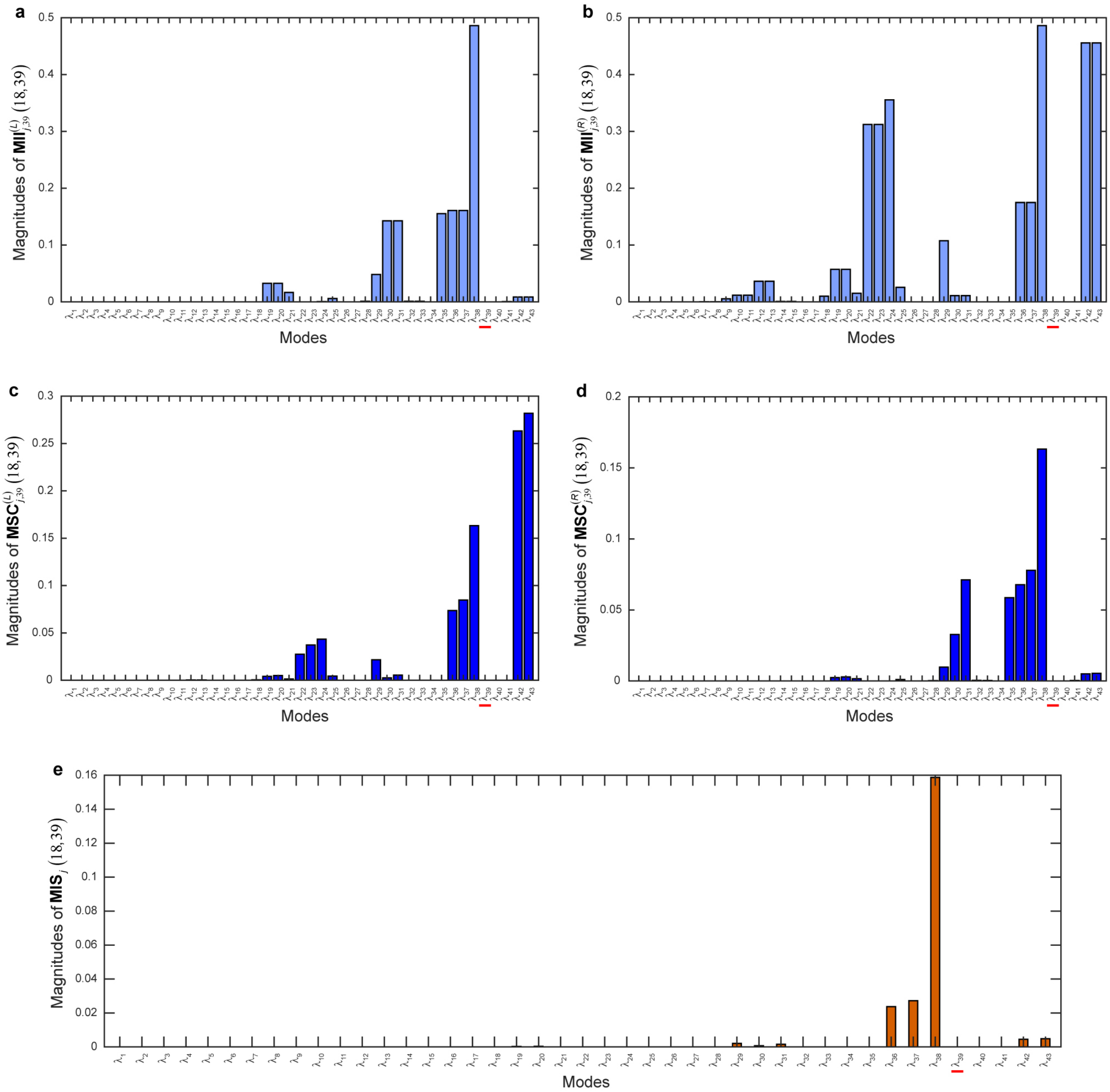

**Fig. 4. MSC, MII, and MIS of $\lambda_{39}$ with respect to $x_{18}$ for the converter-dominated power system. a**, Magnitudes of $\mathbf{MII}^{(L)}_{j,\,39}$ (18, 39). **b**, Magnitudes of $\mathbf{MII}^{(R)}_{j,\,39}$ (18, 39). **c**, Magnitudes of $\mathbf{MSC}^{(L)}_{j,\,39}$ (18, 39). **d**, Magnitudes of $\mathbf{MSC}^{(R)}_{j,\,39}$ (18, 39). **e**, Magnitudes of $\mathbf{MIS}_j$ (18, 39). Several modes (e.g., $\lambda_{35}$, $\lambda_{36}$, $\lambda_{37}$, and $\lambda_{38}$ in **a** and $\lambda_{24}$, $\lambda_{38}$, $\lambda_{42}$, and $\lambda_{43}$ in **b**) exhibit strong interactions with $\lambda_{39}$ in fixed space. However, according to the MSCs in **c** and **d**, the modal interactions of $\lambda_{35}$ and $\lambda_{24}$ are not sensitive to this perturbation. The MIS in **e** is obtained by weighting the MII with the MSC. The MIS with the largest magnitude is observed for $\lambda_{38}$, indicating that this mode dominates the perturbation-driven reconfiguration of the modal interaction with $\lambda_{39}$ in perturbation space. The magnitudes of $\mathbf{MIS}_j$ (18, 39) in $\lambda_{36}$, $\lambda_{37}$, $\lambda_{42}$, and $\lambda_{43}$ are also shown in **e**. Then, $\mathbf{PFS}_{18,\,39}$ is interpreted as the sum of $\mathbf{MIS}_j$ (18, 39). Note that the magnitude of $\mathbf{MIS}_{39}$ (18, 39), i.e., the self-interaction sensitivity of a mode, is zero, which is consistent with the theoretical formulation.

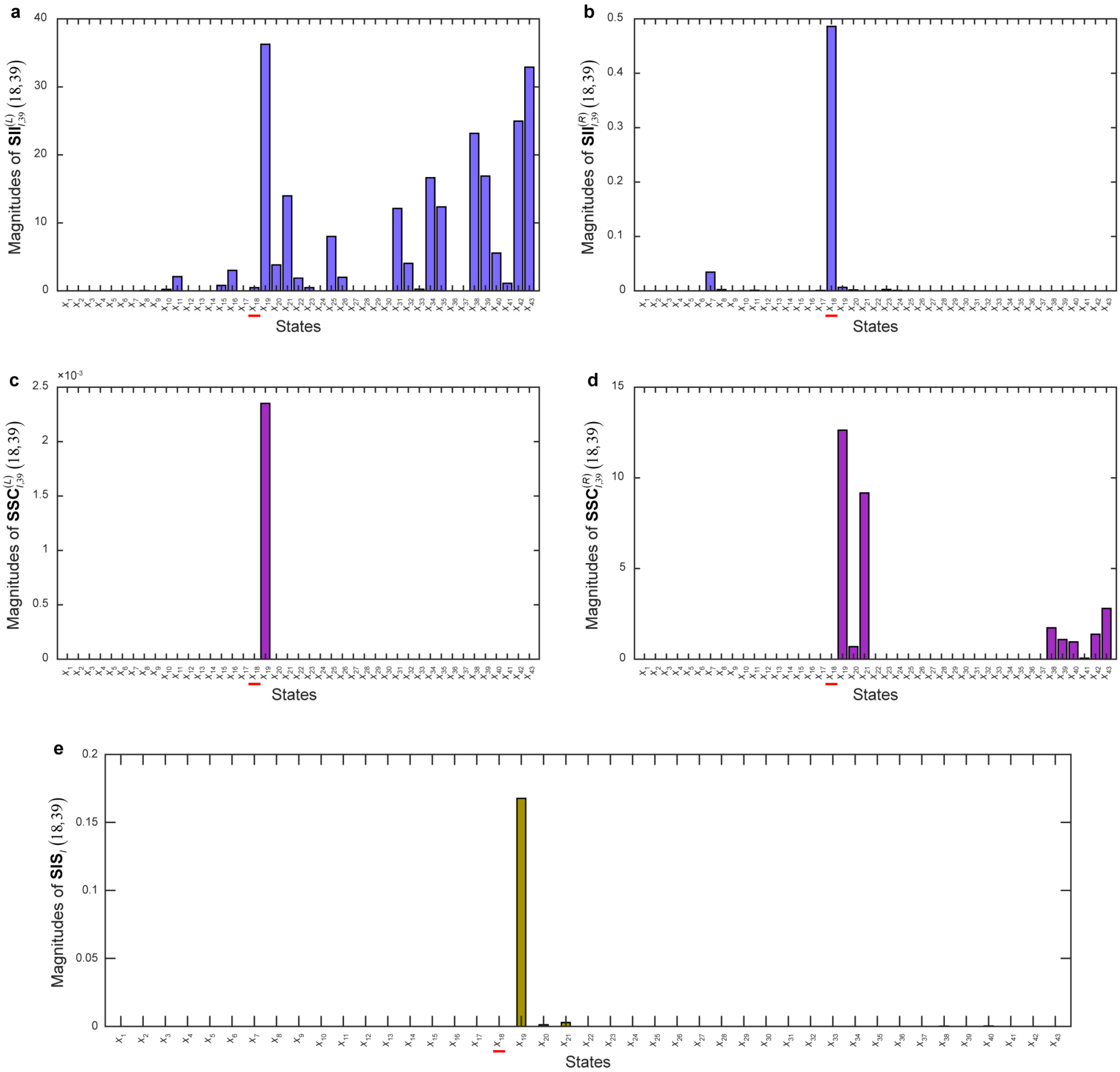

**Fig. 5. SSC, SII, and SIS of $x_{18}$ with respect to $\lambda_{39}$ for the converter-dominated power system.** **a**, Magnitudes of $\mathbf{SII}^{(L)}_{l,\,39}$ (18, 39). **b**, Magnitudes of $\mathbf{SII}^{(R)}_{l,\,39}$ (18, 39). **c**, Magnitudes of $\mathbf{SSC}^{(L)}_{l,\,39}$ (18, 39). **d**, Magnitudes of $\mathbf{SSC}^{(R)}_{l,\,39}$ (18, 39). **e**, Magnitudes of $\mathbf{SIS}_l$ (18, 39). According to **a** and **b**, multiple states exhibit strong interactions with $x_{18}$ in fixed space. In contrast with these results, the SSCs indicate that only a subset of the states responds sensitively to the perturbation. When $\lambda_{39}$ is fixed, the SIS in **e** is obtained by weighting the SII with the SSC. The SIS with the largest magnitude is observed for $x_{19}$, followed by $x_{21}$ and $x_{20}$, indicating that these states dominate the perturbation-driven reconfiguration of the state interactions with $x_{18}$ in perturbation space. Then, $\mathbf{PFS}_{18,\,39}$ is interpreted as the sum of $\mathbf{SIS}_l$ (18, 39). Although the self-interaction sensitivity of $x_{18}$ with respect to $\lambda_{39}$, which is $\mathbf{SIS}_{18}$ (18, 39), is defined by the SIS, its magnitude is nearly zero, i.e., $7.114 \times 10^{-14}$.